\documentstyle[aps,prl,twocolumn]{revtex}

\def\be{\begin{equation}}
\def\bea{\begin{eqnarray}}
\def\bma{\begin{mathletters}}
\def\ee{\end{equation}}
\def\eea{\end{eqnarray}}
\def\ema{\end{mathletters}}

\begin{document}
\author{Vlatko Vedral}
\title{The Meissner effect and massive particles as witnesses of macroscopic entanglement}
\address{The Schr\"odinger Institute for Mathematical Physics, Boltzmanngasse 9, A-1090 Vienna, Austria \\and\\
The School of Physics and Astronomy, University of Leeds, Leeds LS2 9JT, United Kingdom}
\date{\today}
\maketitle

\begin{abstract}
We show how the spatial macroscopic entanglement equivalent to the
off diagonal long range order (ODLRO) implies the Meissner effect
and flux quantisation for a superconductor. It is argued by
analogy with superconductors that the Higgs field must also be
entangled in the same way. Internal (spin) entanglement is shown
to be irrelevant within this context, although it can of course
also be computed.
\end{abstract}

\vspace*{1cm}

\noindent {\bf Introduction}. We have recently argued that high temperature macroscopic
entanglement is possible and linked it to high $T_c$ superconductivity
\cite{Vedralnew}. We have also discussed the relationship between the notion of long range off
diagonal order (ODLRO) \cite{Yang1} in a state and the existence of bipartite and
multipartite entanglement in the same state. Now we intend to extend this line of thought
and show that the said multipartite entanglement implies two typical
superconducting effects: the exclusion of the magnetic field from a simply connected
superconductor (the Meissner effect) and
the quantization of flux in multiply connected regions. Although we will use the Hubbard related
models to aid the discussion in this paper, our results hold for any model which displays ODLRO, i.e. normal
superconductors and superfluids. The paper will be finished with a speculation, maintaining that
if Higgs bosons are found to exist, and if they are responsible for mass generation through
the symmetry breaking mechanism, then they must also be entangled. The reason is that the condensation
of Higgs bosons is understood to be the most likely mechanism for mass generation in local gauge field
theories (i.e. the Standard Model).

\noindent {\bf Setting the scene}. The model we analyse
consists of a number of lattice sites, each of which can be
occupied by fermions of spin up or spin down. Since fermions obey
the Pauli exclusion principle, we can have at most two fermions
attached to one and the same site. Let us introduce fermion
creation and annihilation operators, $c^{\dagger}_{i,s}$ and
$c_{i,s}$ respectively, where the subscript $i$ refers to the
$i$-th lattice site and $s$ refers for the value of the spin,
$\uparrow$ or $\downarrow$. The $c$ operators satisfy the
anticommutation relations: $\{c_{i,s},c^{\dagger}_{j,t}\} =
\delta_{ij}\delta_{s,t}$, and $c$'s and $c^{\dagger}$'s
anticommute as usual. (Some general features of fermionic
entanglement were analysed in \cite{Zanardi,Vedral3,rest}).

We only need assume that our model has the interaction which
favours formation of Cooper pairs of fermions of opposite spin at
each site -- these states are known as $\eta$ states \cite{Yang2}
and will be discussed below. The actual Hamiltonian is not
relevant for our present purposes. Suffice it to say that the $\eta$
states are eigenstates of the Hubbard and realted models
relevant for superconductivity \cite{Yang2,Korepin}.

\noindent {\bf Introducing $\eta$ states.}
Suppose that there are $n$ sites and suppose, further, that we introduce an
operator
\begin{equation}
\eta^{\dagger} = \sum_{i=1}^n c^{\dagger}_{i,\uparrow} c^{\dagger}_{i,\downarrow} \;
\end{equation}
that creates a coherent superposition of a Cooper pair in
each of the lattice sites. This $\eta^{\dagger}$ operator can be applied to the vacuum a number of times,
each time creating a new coherent superposition. However, the number of applications,
$k$, cannot exceed the number of sites, $n$, since we cannot have more than one pair
per site due to the exclusion principle. We now introduce the following basis
\begin{equation}
|k,n-k\rangle := {n \choose k}^{-1/2} (\eta^{\dagger})^k |0\rangle \; ,
\end{equation}
where the factor in front is just the necessary normalisation.
Here, the vacuum state $|0\rangle$ is annihilated by all $c$
operators, $c_{i,s} |0\rangle = 0$. We note that the
originally defined $\eta$ operators can also have phase factors
dependent on the location of the site on the lattice, like so
$\eta_k = \sum_n e^{ikn} c^{\dagger}_{n,\uparrow}
c^{\dagger}_{n,\downarrow}$. All the states generated with any
$\eta_k$ from the vacuum have the same amount of entanglement so
that the extra phases will be ignored in the rest of the paper. However,
the phase must be chosen, for
otherwise, if we average over all possible phases, the resulting
state is no longer entangled. Choosing a phase amounts to
``symmetry breaking" and we will have more to say about it below.

We can think of the $\eta$ states in the following way \cite{Vedralnew}. Suppose that $k=2$. Then this
means that we will be creating two $\eta$-pairs in total, but they cannot be created
in the same lattice site. The state $|2,n-2\rangle$ is therefore a symmetric
superposition of all combinations of creating two pairs at two different sites. Let
us, for the moment, use the label $0$ when the site is unoccupied and $1$ when it is
occupied. Then $|2,n-2\rangle =
(|00...11\rangle + ...|11...000\rangle)/\sqrt{n \choose 2}$,
i.e. the state is an equal superposition of states containing $2$ states $|1\rangle$ and
$n-2$ states $|0\rangle$. These states, due to their high degree of symmetry, are much
easier to handle than general arbitrary superpositions and we can compute entanglement
for them between any number of sites \cite{Vedral1}.
In this description each site
effectively holds one quantum bit, whose $0$ signifies that the site is empty and $1$
signifies that the site is full.

\noindent {\bf ODLRO}. The main characteristic of $\eta$ states is
the existence of ODLRO, which implies its various superconducting
features, such as the Meissner effect and the flux quantisation
\cite{Nieh}. The ODLRO is defined by the off diagonal matrix
elements of the two-site reduced density matrix being finite in
the limit when the distance between the sites diverges. Namely,
\begin{equation}
\lim_{|i-j|\rightarrow \infty} \langle c^{\dagger}_{j,\uparrow}
c^{\dagger}_{j,\downarrow} c_{i,\downarrow} c_{i,\uparrow} \rangle \longrightarrow
\alpha \label{ODLRO}
\end{equation}
where $\alpha$ is a constant (independent of $n$) \cite{Yang1}. We will show that although the
existence of off diagonal matrix elements does not guarantee the existence of
entanglement between the two sites, it does guarantee the existence of multi-site
entanglement between all the sites. Note that here, by ``correlations" we mean
correlations between the number of electrons positioned at different sites $i$ and
$j$. This is different from spin-spin
correlations, which would look at the occurrences of both electron spins being up or
down, or one being up and the other being down \cite{Vedral3}.

The $\eta$ states are always of the form
$|k, n-k\rangle := (\hat{S}|000...11\rangle)/\sqrt{n \choose k}$,
where $\hat{S}$ is the total symmetrisation operator. The reason why $\eta$
states are important for
superconductivity is that this phenomenon can be understood to
arise through a condensation of Cooper pairs. Condensation means
that the temperature is so low that all particles are spread
across the whole system -- i.e. their wavelengths are as large as the
system -- so that all
their wave functions overlap to a high degree (using the language
of the first quantisation). This is why the $\eta$ states are a
good description as they represent equal superpositions of Cooper
pairs across all the
sites.

We would now like to start to
compute the entanglement between every two sites
in the state $|k,n-k\rangle$. A simpler and more insightful task would be first to tell if and
when every two qubits in a totally symmetric states are entangled. For this, we need
only compute the reduced two-qubit density matrix which can be written as:
\begin{equation}
\rho_{12}(k) = a |00\rangle \langle 00| + b |11\rangle \langle 11|
+ c |\psi^{+}\rangle \langle \psi^{+}|
\label{density}
\end{equation}
where $|\psi^{+}\rangle = (|00\rangle + |11\rangle)/\sqrt{2}$ and
$a = \frac{k(k-1)}{n(n-1)}$, $b \frac{(n-k)(n-k-1)}{n(n-1)}$ and
$c = \frac{2k(n-k)}{n(n-1)}$. We can easily check that $a+b+c =1$
and so the state is normalised. This density matrix is the same no
matter how far the two sites are from each other, since the state
is symmetric, and must therefore be identical for all qubits. We
can easily test the Peres-Horodecki (partial transposition)
condition for separability of this state \cite{Vedralnew}. This
leads to states $\rho_{12} (k)$ being entangled if and only if $a
+ b - \sqrt{(a-b)^2 + 4c^2} < 0$, which leads to $(k-1) (n-k-1) <
k (n-k)$. This is, of course, satisfied for all $n\ge 2$ and $1 \le k \le n-1$.
So, apart from the case
when the total state is of the form $|000..0\rangle$ or
$|111..1\rangle$, there is always two-qubit entanglement present
in symmetric states. Note, however, that in the limit of $n$ and
$k$ becoming large -- no matter what their ratio may be -- the
value of the left hand side approaches the value of the right hand
side and entanglement thus disappears in the thermodynamical limit
\cite{Vedralnew}.

\noindent {\bf Macroscopic (spatial) entanglement.} The two point
correlation function used in
the calculation of the ODLRO in eq. (\ref{ODLRO}) is, in fact,
just one of the $15$ numbers we need for the full two-site
density matrix. In our simplified
case of symmetric states in the $\eta$-pairing model, this off
diagonal element is equal to $c$. However, for the density matrix
we still need to know $a$ and $b$, and these numbers clearly
affect the amount of entanglement. So, the first lesson is that
two-site entanglement is not the same as the existence of ODLRO,
and therefore two-site entanglement is not relevant for
superconductivity. This does not mean, of course, that there is no
entanglement in the whole of the lattice. In fact, the ODLRO
implies that the two site density matrix contains classical
correlations. This, together with the fact that the overall state
of all electrons is pure, means that there is always bipartite
entanglement present. For example, entanglement exists in the
thermodynamical limit between two bunches of sites containing $k$
and $n-k$ sites respectively, and each bunch containing a site
from our two site density matrix \cite{Vedralnew}.

\noindent {\bf The Meissner effect}. Suppose that we exchange two electron pairs, one
at the site $1$ and one at the site $n$. We can imagine doing this adiabatically,
although the requirement of adiabaticity is by no means necessary (it is merely convenient, as
this evolution then generates no other effect apart from the one we wish to concentrate on). Suppose
that whatever the total state is, the reduced density matrix of sites $1$ and $n$ has a non-vanishing
component of the state
\begin{equation}
|\Psi \rangle = |0_1\rangle |1_{n}\rangle + |1_1\rangle |0_{n}\rangle \; .
\end{equation}
which, from the above discussion, means that we assume ODLRO.
Then, after the swap, this component will look like
\begin{equation}
|\Psi \rangle = |0_1\rangle |1_{n}\rangle +  e^{i\Phi}|1_1\rangle |0_{n}\rangle \; ,
\end{equation}
where $\Phi = \int A dl$ is the line integral of the vector
potential along the path traversed by the electron pair (with
proper units introduced below). The reason why the two states in
the superposition acquire different phases is that the electron
pairs in two states undergo evolutions in opposite directions of
each other -- this is, in fact, the well known Aharonov-Bohm phase
\cite{AB}. So, if in the first state the pair takes one path (i.e.
the electron pair from site $n$ moves to site $1$), in the second
state it takes the reverse of the same path (i.e. the electron
pair from site $1$ moves to site $n$). The off-diagonal element in
the $|0\rangle, |1\rangle$ basis of the two site density matrix of
sites $1$ and $n$ undergoes the following transformation $c
\rightarrow e^{i\Phi}c$. However, the overall state must be totally
symmetric, and therefore
\begin{equation}
e^{i\Phi} = 1 .
\end{equation}
From this we can conclude that
\begin{equation}
\Phi = \frac{2e}{\hbar c} \int A dl = \frac{2e}{\hbar c} \int\int B dS = 2n\pi .
\label{flux}
\end{equation}
But, in the two (three) dimensional space, the electron pair can
take any trajectory, and the only possible choice that satisfies the above is
$B=0$ and $n=0$. Therefore, in a connected region of a material exhibiting bipartite
entanglement there is no magnetic field present. This is the Meissner effect.
We note that it is well known that the magnetic field does penetrate the superconductor
to a very small degree (falling off exponentially with the distance from the surface), but
this cannot be explained with the present simple formalism and we need a more elaborate 
electrodynamic treatment.

\noindent {\bf Flux Quantisation}. Imagine instead that the region
is not simply connected and that there is a hole in the middle
pierced through by a magnetic field (there could be more than one
hole and the same conclusion will hold for each of them). Then
its flux must be quantised. This follows immediately from eq.
(\ref{flux}), since now not all paths are allowed. Namely, the
field is now confined to the region where electrons cannot go, so
that
\begin{equation}
\frac{2e}{\hbar c} \int\int B dS = \frac{2e}{\hbar c} \Phi_c \neq 0 .
\end{equation}
Therefore, we must have that $\frac{2e}{\hbar c} \Phi_c = 2n \pi$,
and so the flux is quantised in units of $\hbar c/2e$:
\begin{equation}
\Phi_c = n \frac{\hbar c}{2e} .
\end{equation}
This is the flux quantisation effect. Note that the denominator
contains twice the electron charge and this is a consequence of
electrons forming Cooper pairs.

The flux quantisation that we have just derived lies behind the persistent flow
of electrical current in a superconductor. The flux is a consequence of the flowing current and
any (continuous) dissipation cannot change the current continuously as the flux
is discrete. Therefore the current persists indefinitely.

Finally, if there is no ODLRO, meaning that as $n\rightarrow
\infty$ we have that $c\rightarrow 0$ in eq. (\ref{density}), then 
neither of the above two effects follow. Any phase now gained upon
exchange of electrons as described above will not be reflected in
the two site state and therefore we cannot argue that this phase
has to have a special value. Therefore, bipartite entanglement is
necessary for the Meissner effect and flux quantisation. Note briefly that
the converse is not true. Not all entanglement will lead to superconductivity.
For example, look at the state of two sites of the form: $|00\rangle + |11\rangle$. When
we exchange the pairs we get no extra phase in the second ket (because they have opposite
signs and so their product equals identity), so that the state remains the same. Therefore,
any magnetic field is allowed to permeate such a state. This is why the ODLRO concerns the coherences
between states $|01\rangle$ and $|10\rangle$.

\noindent {\bf Mass from entanglement between Higgs bosons.} We
would now like to talk about the Higgs mechanism as the main
explanation for the appearance of mass in local gauge field
theories (see Weinberg \cite{Weinberg} for a comprehensive
introduction). In the modern field theory, gauge invariance of the
Hamiltonian (or Lagrangian, which is more typically used) is
invoked to explain the appearance of fields and their bosonic
mediators. By ``gauge transformation" we mean a transformation
that acts on the wavefunction (or the field, more precisely) in
the following manner $|\Psi (x,t)\rangle \rightarrow e^{\theta
(x,t)} |\Psi (x,t)\rangle$, where $\theta (x,t)$ is just a phase
that dependents both on space and time (i.e. it is local). In
order for the Hamiltonian to remain invariant under this local
change, we need to introduce an extra (vector) field, whose
features exactly cancel out the effects of the local phase change.
The necessary field turns out to be the electromegnetic field and
its bosons are, of course, photons. The important point is that if
we are to derive other forces from local gauge invariance (this
requires phases that are non-commuting -- i.e. matrices, but the
concept is the same), the resulting bosons will always be
massless. This result is intuitively clear: the local phase change
has to be matched between arbitrary points and times and this
therefore requires an infinite range force. The mediators of
infinite range forces have to be massless. So, it appears that
local gauge invariance cannot explain forces whose mediators are
massive. 

A solution to this problem was found by Higgs
\cite{Higgs} (and a number of other people, but Higgs was most
prominent). The idea is that in addition to specifying the
Hamiltonian of the fields, we also need to specify their actual
physical state. This state need not possess the same symmetry
properties of the Hamiltonian (hence the phrase ``symmetry
breaking" \cite{Weinberg}). Suppose now that our local gauge
invariance leads to several interacting massless fields. Suppose
also, that one of the fields -- known as the Higgs field -- has
condensed. In a mechanism that is completely analogous to the
Meissner effect, the other fields will now be ``expelled" from
the Higgs field and will therefore become short range. In other
words, their mediators will become massive. The condensation of
the Higgs field therefore provides a mechanism to maintain local
gauge invariance and have massive gauge fields at the same time,
thereby circumventing previously mentioned limitation (the Higgs
boson also acquires a mass in this process). Whether this is the
correct way of explaining the origin of mass in the Universe is
still unclear as the search for Higgs bosons has so far
been fruitless. However, one conclusion we can draw with more
confidence (following this paper) is that if the Higgs field
exists, then its bosons must be entangled. The reason is that the
ODLRO, necessary for condensation, also implies existence of
entanglement, and this is also true for Higgs condensation. The
(obvious) fact that there are massive objects in the Universe would then
be an entanglement witness of the purely quantum correlations in the
underlying Higgs field.

Here we have to exercise some caution. The entanglement in the Higgs
field is something that we refer to as ``continuous variable entanglement",
as opposed to the discrete degrees of freedom of the Hubbard model above,
and this quantity can become infinite. However, this
is not a serious problem because with enough care this infinity can always
be controlled.

We now show in a very simple example the connection between mass and entanglement
that is meant to substantiate the above dicussion, but is by no means a proof of it.
Suppose we have a massive bosonic free
field, $\phi$, with the usual Lagrangian density
$1/2((\partial_{\mu}\phi)^2 + m^2 \phi^2)$ (this is in $1+1$
dimensions), where $m$ is the (fixed finite) mass. This is an infinite continuous system
and we divide it into two halfs (arbitrarily). Tracing one part out and
computing the von Neuman entorpy of the remaining part results in the entropy of entanglement
of $E \approx \ln 1/m^2 a^2$ \cite{Cardy}, where $a$ is some cutoff used
to avoid the ultraviolet infinity (this divergence may also be avoided by using the relative entropy
with respect to some coarse graining \cite{Casini}, much in the same way as Gibbs did
in classical statistical mechanics). The amount of entanglement clearly
depends on the mass which could therefore be said to witness it.

\noindent {\bf Spin Entanglement}. We would like to finally point
out an interesting curiosity that clarifies the notion of
entanglement we have analysed in this paper. Namely, as we
mentioned before, the relevant entanglement for superconductivity
(and Higgs bosons) is the spatial entanglement between numbers of electrons at
different space points. What about the entanglement between the
spin degrees of freedom? If the electrons occupy the
same site, then they have to be anticorrelated (in the singlet
state) because of Pauli's exclusion principle. If the electrons are
on different sites, then they are not spin correlated in the
$\eta$ state (in the BCS model they would be, for a suffuciently small
distance \cite{Oh}). This is because if we measure an electron in one
site and then in another all four possibilities for their internal
states are equally likely and so there can be no spin correlation
present. There is a limit, however, in which the spin entanglement
becomes relevant. This is when the interaction between sites
dominates the hopping amplitude in the Hubbard model and in the
state where we have one electron per site. Here there is 
no ODLRO and the state is not superconducting. However, the
effective interaction between electrons at neighbouring sites is
now of the Heisenberg type, as electrons can still exchange their
locations and could at the same time have opposite spins.
Therefore, at low temperatures there would be some two site spin
entanglement present in the model, which is albeit not important for
superconductivity.

\noindent {\bf Conclusions}. In one of our previous publications
\cite{Vedralnew} we argued that macroscopic entanglement exists at
high temperatures and is related to high temperature
superconductivity. In the present work we showed that the
consequences of that entanglement are the standard features of
superconductors: the Meissner effect and flux quantisation.
Therefore any experiments confirming these two effects are also
automatically offering evidence for macroscopic entanglement. We
have speculated that if the Higgs mechanism for mass generation is
proven to be correct, then the resulting Higgs bosons will be
found to be entangled. Be that as it may, one question
remains open, both for superconductors, or for any other
more general field. Can we extract this existing entanglement and
use it for information processing? This would be very useful in practice, 
and it would seem that natural macroscopic 
entanglement could offer an infinite amount of quantum
non-locality for genuine quantum information processing. This is
the subject of an ongoing research.

\noindent {\bf Acknowledgements}. We thank the European Commission
and Elsag Spa for financial support.

\end{document}